# In a tight spot, spin and charge separate

Gregory A. Fiete
Department of Physics, The University of Texas at Austin, Austin, TX 78712, USA

*Photons (bosons) confined in a hollow waveguide containing an atomic gas could show spin-charge separation, which is more commonly associated with one-dimensional fermions.*

Subject Areas: **Atomic and Molecular Physics, Optics, Interdisciplinary Physics**

Strange things happen when there is a shortage of space. The Pauli exclusion principle is well known for its effect on fermions, but bosons (for instance, photons) also behave interestingly when pressed for space. For this reason, researchers are very interested in schemes that create this kind of environment, in search of new physics and new technologies.

One of these schemes is outlined by Dimitris Angelakis from the Technical University of Crete, Greece, together with collaborators in Singapore [1]. Writing in Physical Review Letters, they propose an intriguing theoretical scenario in which photons in a spatially confined nonlinear optical medium take on the characteristics of a strongly correlated one-dimensional system of interacting fermions, mimicking a Luttinger liquid [2]. In the Luttinger liquid regime, which can occur for fermions in one-dimensional systems, excitations of the fermions take on a bosonic character. Here, the scenario is essentially the reverse—bosons take on a fermionlike character.

The scheme relies on a one-dimensional hollow fiber acting as a waveguide, into which two atomic species are loaded and subjected to multiple laser fields [3, 4]. Exploiting the method of electromagnetically induced transparency (a technique that renders the medium transparent over a narrow frequency range and leads to extreme dispersion) [5], different laser frequencies are tuned to create strongly interacting light-matter excitations known as polaritons [6]. Angelakis et al. have shown that polaritons can mimic highly tunable (variable interaction strength) Luttinger liquid behavior in the regime of effective repulsive interactions. This regime is experimentally accessible for realistic conditions. In order to appreciate this result of correlated fermionlike behavior in a gas of strongly interacting photons, it is helpful to first review some fundamentals of Fermi gases.

Fermi gases are known to have important properties that distinguish them from their bosonic counterparts. Rather than piling into the lowest energy state at low energies and condensing (as bosons do), fermions are forbidden by the Pauli exclusion principle to doubly occupy any quantum state. As a result, fermions "stack up" on top of each other in energy space. In a translationally invariant state at zero temperature, a "Fermi sea" results, in which all momentum states below a certain energy (the Fermi energy) are occupied, while all states above it remain empty. The boundary between the occupied and unoccupied states in momentum space is referred to as the Fermi surface. States deep in the Fermi sea generally do not have transitions between them, because all such states are occupied. Thus the low-energy excitations of a many-body Fermi system are dominated by the properties near the Fermi surface. In one of the crowning achievements of 20th century physics, it was shown that this basic picture for fermions remains true even in the presence of interactions of the order of the kinetic energy. In particular, low-energy excitations in the interacting system evolve smoothly from the noninteracting fermions and possess the same spin and charge. The so-called Fermi-liquid theory describes such a system and finds important applications in He-3 and many metals. It is believed to capture the "universal" low energy properties of Fermi gases, provided, of course, that other phases such as superconductivity do not intervene.

However, there is a hitch: Fermi-liquid theory breaks down in one spatial dimension due to the rather singular scattering that occurs between the two Fermi wave vectors corresponding to right and left moving fermions at the Fermi energy. Scattering across the Fermi surface (which in one spatial dimensional is two points in momentum space) has a significant consequence and drives the system towards a different low-energy phase, the Luttinger liquid [2]. A remarkable feature of a Luttingerliquid is the nature of its low-energy excitations, which are collective bosonic excitations rather than the single-particle fermionic excitations described by Fermiliquid theory. Moreover, under rather general conditions, the Luttinger liquid exhibits spin-charge separation, in which the spin and charge of the fermions possess independent dynamics, so the velocities of the spin and charge excitations of a Luttinger liquid are distinct. For a familiar fermion like the electron, this surprising feature runs counter to the usual intuition that both charge and spin are "tied" to a particle, and it has the remarkable consequence of spin and charge disturbances that propagate away at different speeds from an electron suddenly injected into a Luttinger liquid.

While the unusual properties of one-dimensional Fermi systems have been studied theoretically for decades [2], their experimental identification has proven more challenging, with significant advances coming only in the last decade. A compelling identification of a Luttinger liquid in experiment can only be made if one has access to its dynamical properties, described by the spectral function $A(q,w)$, which contains information about the energy-momentum relationships of the excitations. This quantity allows for an unambiguous measure of the different spin and charge excitations predicted by theory. A direct measurement of the spectral function for a spin-charge-separated Luttinger liquid system was achieved by Auslaender et al. in 2005 in a parallel quantum wire geometry in a semiconducting heterostructure [7]. In spite of this accomplishment, experimental Luttinger liquid systems with a wide range of tunability and operating conditions remain rare.

Thus the suggestion by Angelakis et al.[1] that spincharge- separated Luttinger liquid behavior could occur for photons in a one-dimensional nonlinear medium opens the door to a host of potential new studies and applications of spin-charge separation in one dimension. The basic idea is to employ a one-dimensional nonlinear optical media with two species of atoms to create a gas of strongly interacting polaritons of two "types," which will become the analogs of spin and charge. The proposal draws on earlier work by Chang et al.[6] in which it was shown that a regime of very strongly interacting polaritions could be achieved in a single-component one-dimensional nonlinear optical media. In the singlecomponent case, the strongly interacting polariton system was shown to realize a Tonks gas, a strongly interacting system of bosons with contact (zero range) interactions [6], which has already been observed in cold atomic gases (as opposed to photons) of bosons [8, 9]. When the interactions in a system of bosons are very strong, particles tend to avoid the same spatial location, which mimics the Pauli exclusion principle for fermions. Of this effect, one sometimes says that the strong interactions have caused the boson system to "fermionize." Once a one-dimensional bosonic system reaches this regime, the connection to interacting fermions becomes clear, and the mathematical descriptions of the low-energy behavior are identical [1, 2, 6].

FIG. 1: **(See online version for figure)** (a) Schematic of the device. Two different species (yellow and blue) populate the waveguide. Two counterpropagating quantum light fields (E) and two classical control beams (W) create spin and charge separation, represented by the red and green wavefronts above the waveguide. (b) Energy level diagrams for the two species of atom. Four atomic levels are exploited for optimal tunablility. The strongly interacting polariton gas is created by shifting in the fourth level after a pulse of photons enters the fiber. (Credit: Alan Stonebraker, adapted from D. Angelakis et al.[1])

A schematic of the experimental proposal is shown in Fig. 1. After loading two atomic species into the hollow core fiber [Fig. 1 (a)], two counter-propagating control beams W1,2 (for each atomic species) establish a standing wave inside the waveguide, forming a Bragg grating that traps the quantum fields E1,2 inside the medium. Four atomic levels for each species are exploited for optimal tunability [Fig. 1(b)]. The strongly interacting polariton gas is created by shifting in the fourth level after a pulse of photons enters the fiber. (Before tuning in the fourth energy level, the polariton gas is only weakly interacting.) Once the desired state is established, one of the control beams, say W1+, can be turned off, resulting in a spin-charge-separated pulse that moves towards the end of the fiber. The resulting evolution of the pulse, which is mostly photon in character, will exhibit spincharge separation by moving with two distinct velocities.

The authors propose two methods to detect this signal. (i) Measure the time evolution of a single excitation, similar to that done in cold atom systems, by probing the correlations in the intensity of light emitted from the fiber. From this information, one may infer spin and charge densities and their corresponding velocities. (ii) Directly measure correlations established during evolution of the Hamiltonian via the single-particle spectral function (as was measured in [7] for electrons). Numerical estimates for experimental parameters required to reach the regime of strong interactions and to observe the spin-charge separation suggest it is within current technical capability.

The strength of a proposal like that of Angelakis et al.[1] is that it highlights the universality of physical phenomena that might seem at first to be restricted to one particular set of conditions. Whenever such new scenarios are uncovered, they inevitably bring some novel aspects to the phenomenon. In this case, there is a high degree of tunablity and a more readily accessible time domain. If a technological application of onedimensional spin-charge separation is envisioned, polariton systems may lead the way. A tantalizing possibility is to make use of this system as an optical analog of some electronic Hamiltonian that cannot be solved by any conventional means. Employing the polariton system as a "quantum computer" for the electronic scenario, we would then be able to study the correlated fermion problem through this analogous optical system. Additionally, designing polariton scenarios that mimic finite temperature as well as spin-charge separation [10] would allow one to access interesting one-dimensional regimes like the spin-incoherent Luttinger liquid [11] or spin-incoherent-like behaviors in models effectively at zero temperature [12]. The answers to such theoretical questions and their experimental study are likely not far off.